\documentclass[aps,prb,twocolumn,superscriptaddress,amsmath]{revtex4-2}
\usepackage{graphicx}
\usepackage{hyperref}
\usepackage{mathrsfs}
\usepackage{bm}
\usepackage{color}
\usepackage{siunitx}
\hypersetup{hypertex=true,
	colorlinks=true,
	anchorcolor=blue,
	linkcolor=blue,
    citecolor=blue}
\begin{document}

\title{Dynamical theory of chiral-induced spin selectivity in electron donor--chiral molecule--acceptor systems}

\author{Tian-Yi Zhang}
\affiliation{International Center for Quantum Materials, School of Physics, Peking University, Beijing 100871, China}

\author{Yue Mao}
\affiliation{International Center for Quantum Materials, School of Physics, Peking University, Beijing 100871, China}

\author{Ai-Min Guo}
\affiliation{Hunan Key Laboratory for Super-microstructure and Ultrafast Process, School of Physics, Central South University, Changsha 410083, China
}

\author{Qing-Feng Sun}
 \email[]{sunqf@pku.edu.cn}
\affiliation{International Center for Quantum Materials, School of Physics, Peking University, Beijing 100871, China}
\affiliation{Hefei National Laboratory, Hefei 230088, China
}

\begin{abstract}
The chiral-induced spin selectivity (CISS) effect, a phenomenon where the chirality of molecules imparts significant spin selectivity to electron transfer processes, has garnered increasing interest among the chemistry, biology, and physics communities. Although this effect was discovered more than a decade ago, the dynamical process of how electron spin polarization is caused by chiral molecules is still unclear. Here, we propose a dynamical theory of electron transfer in donor--chiral molecule bridge--acceptor systems without electrodes or substrates based on the Lindblad-type master equation. We demonstrate that the molecular spin-orbit coupling generates unequal spin velocities and achieves steady spin polarization with the help of dephasing. Our work elucidates the dynamical process of CISS and may promote the applications of chiral-based spintronic devices.
\end{abstract}

\maketitle

\section{\label{SEC1} Introduction}

Molecular chirality has drawn widespread attention for its vital role in chemical reactions and biological processes \cite{Bornscheuer2012}. Recent research has focused on the interplay between molecular chirality and electron spin, where chiral molecules act as spin filters, converting the traversing electrons from spin-unpolarized to highly spin-polarized \cite{Naaman2015,Naaman2020,Naaman2019}. This interesting phenomenon is called chiral-induced spin selectivity (CISS) effect and it was first observed in stearoyl lysine photoemission experiments \cite{Ray1999}. Since its discovery, CISS has been manifested in various molecular materials \cite{Gohler2011,Mishra2013,Mondal2015,Kumar2017,Abendroth2019,Huizi-Rayo2020,Lu2020,Kim2021,Al-Bustami2022,Aizawa2023,Adhikari2023} and the filtering effect persists even at room temperature \cite{Kim2021,Dor2017,Zhang2023,Xu2023}. To explain the origin of CISS, many theories have been proposed successively \cite{Guo2012,Guo2012_2,addr1,Guo2014_1,Guo2014,Pan2015,Wang2024,Gersten2013,Alwan2021,Liu2021,Monti2024,Medina2012,Nurenberg2019,Fransson2020,Zhang2020,Das2022,Fransson2023,Vittmann2023,Fransson2019,Chiesa2024,addr2,addr3}, including the chiral molecule model with spin-orbit coupling (SOC) \cite{Guo2012,Guo2012_2,addr1,Guo2014_1,Guo2014,Pan2015,Wang2024}, angular momentum selection \cite{Gersten2013,Alwan2021,Liu2021,Monti2024}, spin-dependent scattering \cite{Medina2012,Nurenberg2019}, electron-phonon coupling \cite{Fransson2020,Zhang2020,Das2022,Fransson2023,Vittmann2023}, and electron correlation \cite{Fransson2019,Chiesa2024}. The spin selection has also been investigated through analytical modeling \cite{Varela2016,Ghazaryan2020,Varela2023} and first-principles calculations \cite{Maslyuk2018,Zollner2020,Dianat2020,Naskar2023}. However, the dynamical process of how non-polarized electrons transform into spin-polarized states when passing through chiral molecules is still unclear, even though it is an essential process in producing CISS.

Recently, CISS has been found in isolated covalent donor--chiral molecule bridge--acceptor (D-B-A) systems using time-resolved electron paramagnetic resonance (EPR) spectroscopy \cite{Eckvahl2023}. The system consists of a donor D, an acceptor A, and a chiral molecule B bridging them in the middle. In the experiment, the electrons are initially in D and spin unpolarized. After laser-induced photoexcitation, electrons in D undergo a series of electron transfer processes through B to A on an ultrafast timescale. A final spin polarization in A can be detected through transient EPR spectra. This experiment provides direct evidence of the CISS effect on the spin dynamics in the ultrafast electron transfer through D-B-A systems without electrodes or substrates. The experimental phenomenon can be partially explained via numerical results from Chiesa \textit{et al.} \cite{Chiesa2024}. However, the underlying ultrafast spin dynamical processes and the fundamental cause of spin polarization are still unclear and require further exploration.

\begin{figure}
	\includegraphics[width=1\columnwidth]{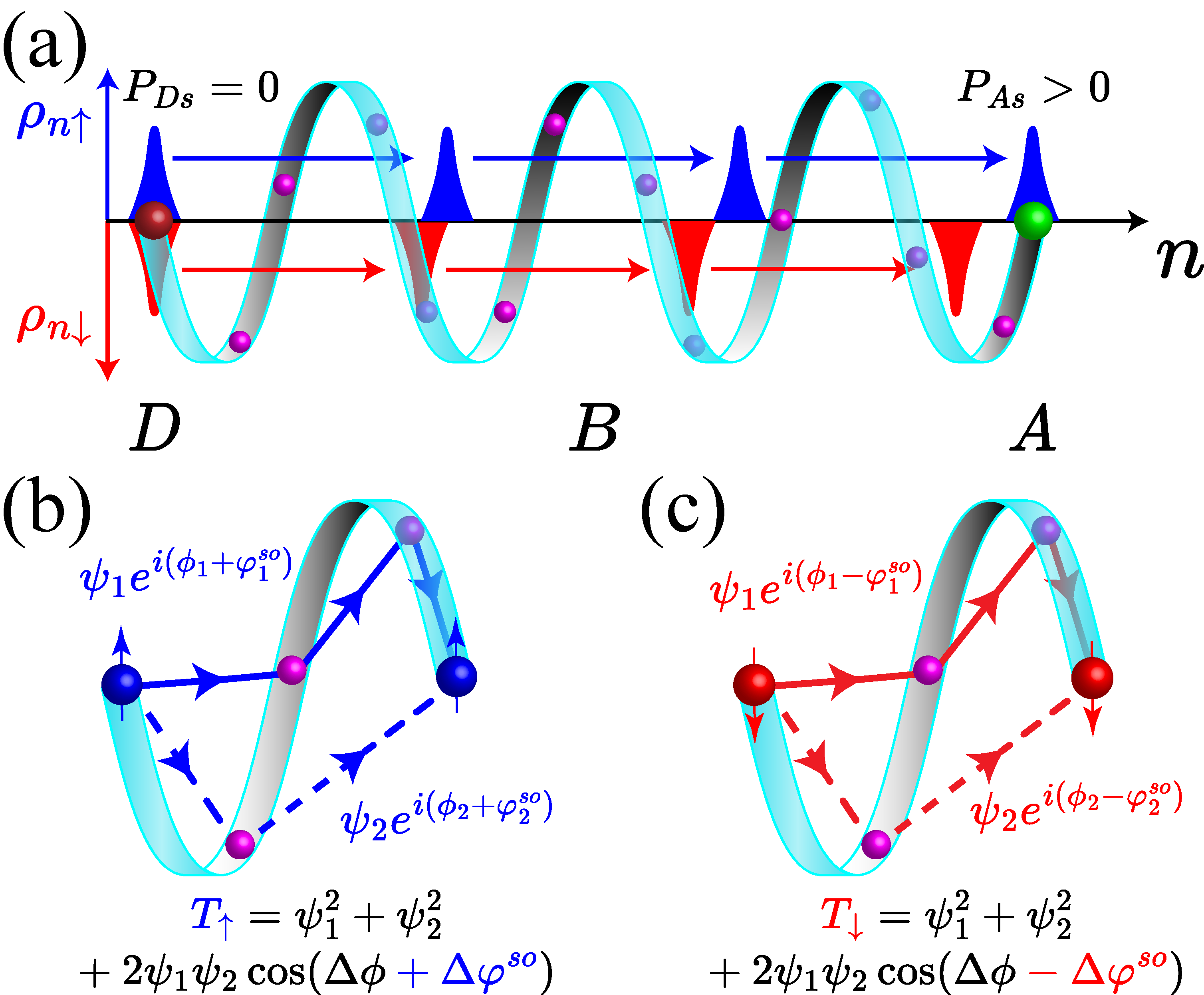}%
	\caption{\label{fig1} (a) Schematics of different spin velocities and spin separation. The transfer starts at D, where the density of spin-up (blue) and spin-down (red) are equal, resulting in zero spin polarization. During the propagation, unequal spin velocity caused by SOC and helical structure induces the separation of different spins. Consequently, spin-up electrons arrive earlier at A, leading to an instantaneous polarization at A. (b) and (c) In the first-order tunneling process, the transmission probability of spin-up electron (b) and spin-down electron (c) are different due to the different signs of phases $\varphi_{1/2}^{so}$ introduced by SOC. The different transmission probabilities lead to different propagation velocities.}
\end{figure}

In this paper, we study the dynamical process of electron spin polarization as they pass through chiral molecules. By calculating the spin dynamics in a D-B-A model [Fig. \ref{fig1}(a)] using a Lindblad-type master equation, we demonstrate that CISS originates from the small SOC and helical structure of chiral molecules, which alters the propagation velocities of different spins. Electrons with a certain spin direction move faster and will arrive at acceptor A first, resulting in instantaneous high spin polarization. Then we focus on the dephasing process of electron propagation, which leads to a conversion from slow spin to fast spin and causes a long-term steady spin polarization. Our theory details how unpolarized spins transform into polarized spins, clarifies the origin of CISS from a dynamical perspective, and may facilitate the manufacture of chiral-based spintronic devices.

The rest of the paper is organized as follows. In Sec. \ref{SEC2}, we introduce the model Hamiltonian of D-B-A systems with B being protein-like single-helical molecules.
In Sec. \ref{SEC3}, we present the physical picture of CISS, demonstrating that the spin polarization is a result of unequal spin velocities influenced by SOC and chiral structure.  In Sec. \ref{SEC4}, we introduce a Lindblad-type master equation and investigate the spin dynamics in D-B-A systems. In Sec. \ref{SEC5}, we study and discuss the effect of the dephasing process. 
Finally, some discussions and a summary are given in Sec. \ref{SEC6}. 
In the main text, we take protein-like single-helical molecules as an example.
In the Appendix, the D-B-A system with double-stranded DNA (dsDNA) as the chiral bridge B 
is studied to validate the robustness and general applicability of our theoretical framework.

\section{\label{SEC2} Model of the D-B-A system}

The spin transport along the D-B-A system can be simulated by the Hamiltonian $H=H_B+H_{BD}+H_A$. This Hamiltonian consists of a donor D, an acceptor A, and a chiral bridge B bridging them in the middle [see Fig. \ref{fig1}(a)] \cite{Luo2021,Chiesa2021,Fay2021,Fay_Limmer2021,Fay2023,Macaluso2023,Chiesa2024}. The first term $H_B$ is the Hamiltonian of the chiral molecule bridge \cite{Guo2014}:
\begin{equation}
    \begin{aligned}
    H_B=&\left(\sum_{n=1}^{N}\varepsilon_n c_{B,n}^\dagger c_{B,n}+\sum_{n=1}^{N-1}\sum_{j=1}^{N-n}[t_j c_{B,n}^\dagger c_{B,n+j}+h.c.]  \right)\\
    &+\sum_{n=1}^{N-1}\sum_{j=1}^{N-n}[2i s_j \cos(\varphi_{n,j}^{-})c_{B,n}^\dagger \sigma_{nj} c_{B,n+j}+h.c.],
     \label{HB}
     \end{aligned}
\end{equation}
where $c_{B,n}^\dagger=(c_{B,n\uparrow}^\dagger,c_{B,n\downarrow}^\dagger)$ is the creation operator at site $n$.
Here, let's take protein-like single-helical molecules as the chiral bridge B as an example.
Another example where dsDNA are taken as the chiral bridge B 
is investigated in the Appendix.
The molecular length is $N$. $\varepsilon_n$ is the on-site energy, $t_j=t_1 e^{-(l_j-l_1)/l_c}$ is the hopping integral between sites $n$ and $n+j$. $l_c$ is the decay exponent. $s_j=s_1 e^{-(l_j-l_1)/l_c}$ is the SOC, $l_j=\sqrt{[2 R \sin(j \Delta \varphi/2)]^2+(j \Delta h)^2}$ is the Euclidean distance between sites $n$ and $n+j$, which corresponds to the helical structure of the chiral protein with radius $R$. $\Delta \varphi$ and $\Delta h$ are the twist angle and the stacking distance between two neighboring sites, respectively. $\sigma_{nj}=[\sigma_x \sin(\varphi_{n,j}^+)-\sigma_y \cos(\varphi_{n,j}^+)]\sin \theta_j + \sigma_z \cos \theta_j $. $\sigma_{x,y,z}$ are the Pauli matrices and $\varphi_{n,j}^{\pm}=(\varphi_{n+j} \pm \varphi_n)/2$, $\varphi_n = n \Delta \varphi$ is the cylindrical coordinate of site $n$.
$\theta_j=\arccos[2 R \sin(j \Delta \varphi/2)/l_j]$ is the space angle.
The second term $H_{BD}$ in $H$ is the Hamiltonian of donor D and the hopping between D and B: $H_{BD}=\varepsilon_D c_D^\dagger c_D +(t_{BD} c_{B,1}^\dagger c_D +h.c.)$, where $c_D^\dagger=(c_{D\uparrow}^\dagger, c_{D\downarrow}^\dagger)$ is the creation operator of D. $\varepsilon_D$ is the on-site energy of D and $t_{BD}$ is the hopping integral between D and B. The last term $H_A$ in $H$ is the Hamiltonian of acceptor A: $H_A=\varepsilon_A c_A^\dagger c_A$, where $c_A^\dagger=(c_{A\uparrow}^\dagger, c_{A\downarrow}^\dagger)$ is the creation operator of A, and $\varepsilon_A$ is the on-site energy of A. The full Hamiltonian
$H$ describes a D-B-A system with a single-helical bridge B (featuring SOC and a multi-pathway structure) between D at the leftmost end and A at the rightmost end. In the experiment, electron dynamics may also be influenced by electron-electron Coulomb interactions. However, since the electrons participating in the transfer process are located on the Fermi surface, the Coulomb interaction from electrons below the Fermi surface manifests as a renormalization of the on-site energy (i.e., $ \varepsilon_n \rightarrow \varepsilon_n + U $, where $ U $ is the Coulomb interaction energy), which does not fundamentally alter the Hamiltonian here. Furthermore, strong Coulomb interactions may lead to unequal electron densities for different spins, thereby inducing magnetization. Nevertheless,
in experiments, chiral molecules are composed of light elements (e.g., C, H, O, N), which are nonmagnetic,
suggesting that the Coulomb interaction has a relatively weak effect on spin polarization. In the following we will show that the SOC and helical structure of B significantly affect the spin-dependent electron dynamics, ultimately leading to the spin polarization observed in experiments.

\section{\label{SEC3} Physical picture of SOC-induced unequal spin velocities}

We then elucidate the physical picture of the interplay between SOC of chiral molecule and the spin of electrons propagating through it. As indicated by the long-range hopping terms ($c_{B,n}^\dagger c_{B,n+j}$) in Eq. (\ref{HB}), there are multiple pathways for electrons to propagate, and we consider two of them as an example, as shown in Figs. \ref{fig1}(b, c). The transmission probability is proportional to
$T_s=|\psi_1 e^{i(\phi_1 + s \varphi_1^{so})}+\psi_2 e^{i(\phi_2 + s \varphi_2^{so})}|^2=\psi_1^2+\psi_2^2+2\psi_1 \psi_2 \cos(\Delta \phi + s \Delta \varphi^{so})$, where $\Delta \phi = \phi_1 - \phi_2$, $\Delta \varphi^{so} = \varphi_1^{so} - \varphi_2^{so}$ with $\psi_1$ and $\psi_2$ the wave-functions, $\phi_1$ and $\phi_2$ the dynamical phases and $\varphi_1^{so}$ and $\varphi_2^{so}$ the spin-related phases through pathways 1 and 2 \cite{Sun2005,Pan2016}. Here $s=\uparrow (+)$ and $\downarrow (-)$ correspond to the spin-up [Fig. \ref{fig1}(b)] and spin-down [Fig. \ref{fig1}(c)] electrons, respectively. As a result, the transmission probabilities for the spin-up and spin-down electrons are usually different, $T_{\uparrow} \neq T_{\downarrow}$. Electrons of one spin have a higher transmission probability and thus have a higher propagation velocity. On the other hand, electrons of the opposite spin have a lower transmission probability, causing a lower propagation velocity.

The unequal propagation velocity of different spins is the key to CISS. Fig. \ref{fig1}(a) illustrates the physical picture of the electron transfer process. The excited unpolarized electrons are initially situated at D and then transferred to A through B. Let $\rho_{n\uparrow}$ and $\rho_{n\downarrow}$ represent the spin density of spin-up and spin-down at position $n$ with $n=D, 1, 2, \cdots N, A$; the local spin polarization is then defined as \cite{Tedrow1971} $P_{ns}=(\rho_{n\uparrow}-\rho_{n\downarrow})/(\rho_{n\uparrow}+\rho_{n\downarrow})$. At the time $t=0$, the number of spin-up electrons at D is the same as that of spin-down ones, resulting in $P_{Ds}=0$. In the transfer process, the different spins gradually separate due to the unequal velocity [see Fig. \ref{fig1}(a)]. Assuming the spin-up electrons move faster, they will outpace the spin-down electrons, generating local polarization. After a certain time of propagation, the spin-up electrons reach A first and start to accumulate, while the spin-down electrons are still on the way to A. This results in an instantaneous spin polarization $P_{As}>0$ in A, which will keep increasing until the arrival of spin-down electrons. During the transfer process, the number of spin-up and spin-down electrons are the same under the restriction of current conservation and the time-reversal invariance \cite{Buttiker1986,Hackenbroich2001}. Therefore, finally, the number of spin-down electrons arriving at A is the same as that of spin-up electrons, causing the spin polarization in A to decrease and return to $P_{As}=0$.

\section{\label{SEC4} CISS dynamical master equation}

We next numerically study the spin dynamical process of the CISS effect. The dynamical process of spin transfer through the D-B-A system can be described by the Lindblad-type master equation \cite{Macaluso2023,Haberkorn1976,Breuer2007}:
\begin{equation}
    \begin{aligned}
        \hbar \frac{d \rho}{d t}=&-i [H,\rho]+\Gamma \left(L_{AB}\rho L_{AB}^\dagger -\frac{1}{2} \{L_{AB}^\dagger L_{AB}, \rho\}\right)\\
        &+\Gamma_d \sum_{n=1}^{N} \sum_{\mu = x,y,z}\left(L_{n\mu} \rho L_{n\mu}^\dagger -\frac{1}{2} \{L_{n\mu}^\dagger L_{n\mu}, \rho\}\right),
        \label{ME}
    \end{aligned}
\end{equation}
where $\rho$ and $H$ are the density matrix and the Hamiltonian of the D-B-A system described in Sec. \ref{SEC2}. The first term on the right-hand side describes the unitary evolution. The second term describes the reflectionless process of electrons entering A from the rightmost end of B, where $L_{AB}=c_A^\dagger c_{B,N}$ is the quantum jump operator and $\Gamma$ is the jumping strength. The last term describes the dephasing process when electrons traveling in B, which is caused by inelastic scatterings from the electrons, the impurities, the nuclear spins, or the phonons \cite{Xing2008,Jiang2009}. This phase-breaking process is revealed in previous works \cite{Morita2003,Skourtis2005,Giese2008}. $L_{n\mu}=c_{B,n}^\dagger \sigma_\mu c_{B,n}$ are operators describing the dephasing process, and $\Gamma_d$ represents the dephasing strength. The electrons are initially localized in D and in a mixed state $\rho(0)=(|D \uparrow \rangle \langle D \uparrow |+|D \downarrow \rangle \langle D \downarrow |)/2$, indicating the number of spin-up and spin-down electrons are equal to each other ($\rho_{D\uparrow}=\rho_{D\downarrow}=0.5$), and the spin polarization of D is $P_{Ds}=0$. As time evolves, electrons will propagate through B and then enter A without returning.

For the D-B-A system, the structural parameters for B are the radius $R = 0.25 \ \rm{nm}$, the twist angle $\Delta \varphi=5\pi/9$, and the stacking distance $\Delta h=0.15 \ \rm{nm}$. The length of B is $N = 30$, and the decay exponent is $l_c=0.9$ \r{A}. The on-site energy is set to $\varepsilon_n = 0$ without loss of generality, and the nearest-neighbor hopping $t_1$ is taken as the energy unit. These parameters are the same as those in the previous work \cite{Guo2014}. The nearest-neighbor SOC is chosen as $s_1=0.03 t_1$. The on-site energy of D is set as $\varepsilon_D=-0.22 t_1$ and the hopping between D and B is $t_{BD}=0.1 t_1$. The on-site energy of A is set as $\varepsilon_A=-0.5 t_1$ and the jumping strength is $\Gamma = t_1$. By setting $t_1 = 0.1 \ \rm{eV}$, we get a tiny SOC $s_1 = 3 \ \rm{meV}$. 
Although the SOC in flat carbon systems (like graphene) is very small, on the order of microelectronvolts \cite{Min2006}, the SOC in curved systems (like carbon nanotubes or chiral molecules) is significantly greater due to the hopping between $\pi$ and $\sigma$ bands \cite{Varela2016,Naaman2020,Huertas-Hernando2006,Salazar2018,Torres2020}.
Experimental measurements have also demonstrated greatly enhanced SOC strengths in curved systems, comparable to those in isolated carbon atoms, with values reaching a few meV \cite{Ghazaryan2020,Steele2013,Jhang2010}. Therefore, our choice of $s_1=3 \ \rm{meV}$ for the molecular SOC parameter aligns well with both theoretical predictions and experimental observations in comparable curved carbon-based systems.

The parameters adopted here result in a time scale of electron transfer about a few ps, correctly describing the ultrafast spin transfer process (see the numerical results below) \cite{Chiesa2024}. The values of all above mentioned parameters will be used throughout the paper except for specific indication in the figure.

\begin{figure}
	\includegraphics[width=1\columnwidth]{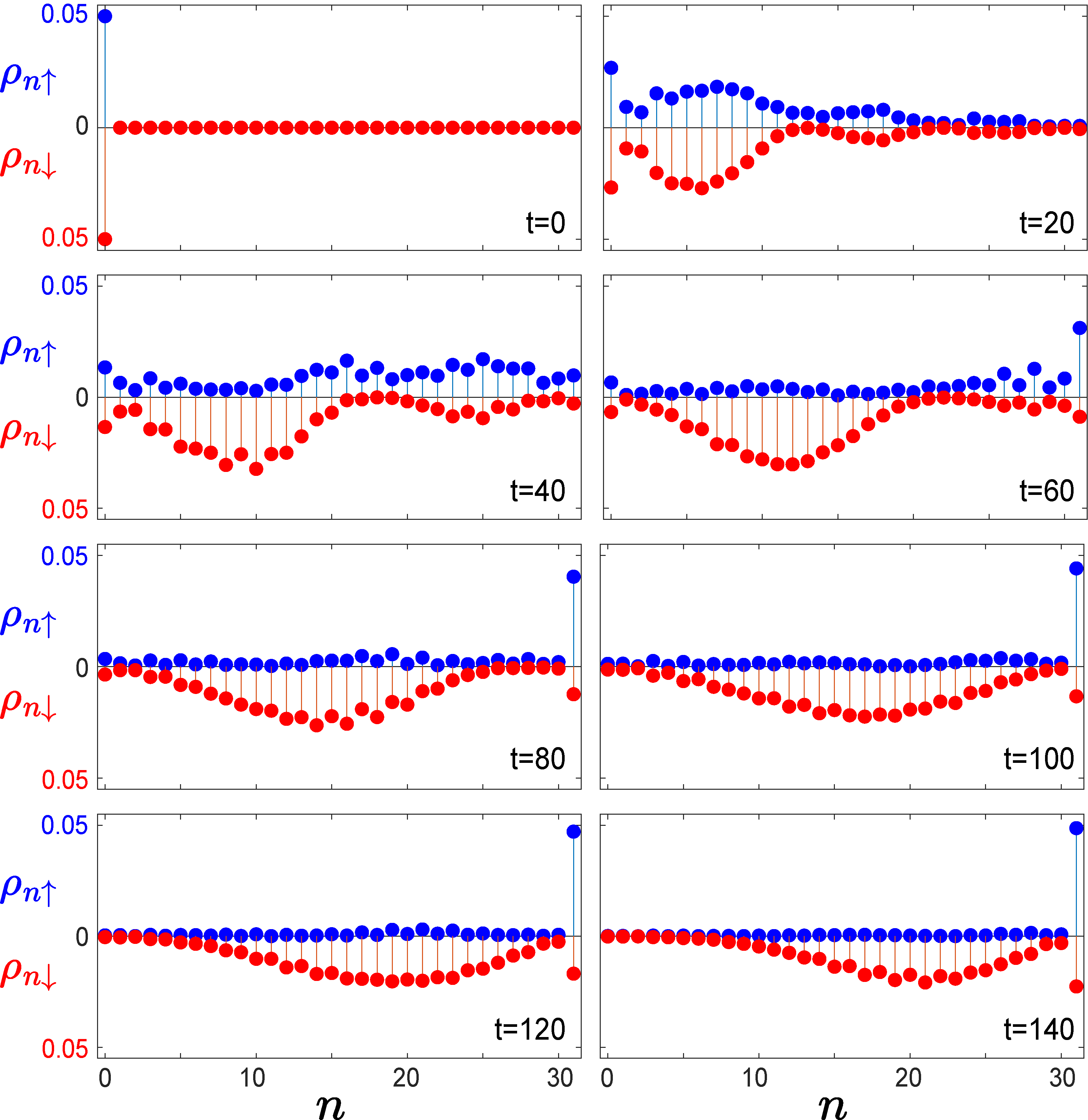}%
	\caption{\label{fig2} Dynamical process of spin transfer without dephasing. Due to the unequal velocity, electrons with spin-up quickly propagate through B and accumulate in A, while electrons with spin-down are still moving in B. The spin density at D ($n=0$) and A ($n=N+1=31$) has been reduced to one-tenth of its original value for better illustration, and the time $t$ labeled in the figures is in the unit of $\hbar/t_1$. The length of B is $N=30$, the SOC is $s_1=0.03t_1$, and dephasing strength is $\Gamma_d=0$. Other parameters are $t_{BD}=0.1t_1$, $\Gamma=t_1$.}
\end{figure}

\begin{figure}
	\includegraphics[width=1\columnwidth]{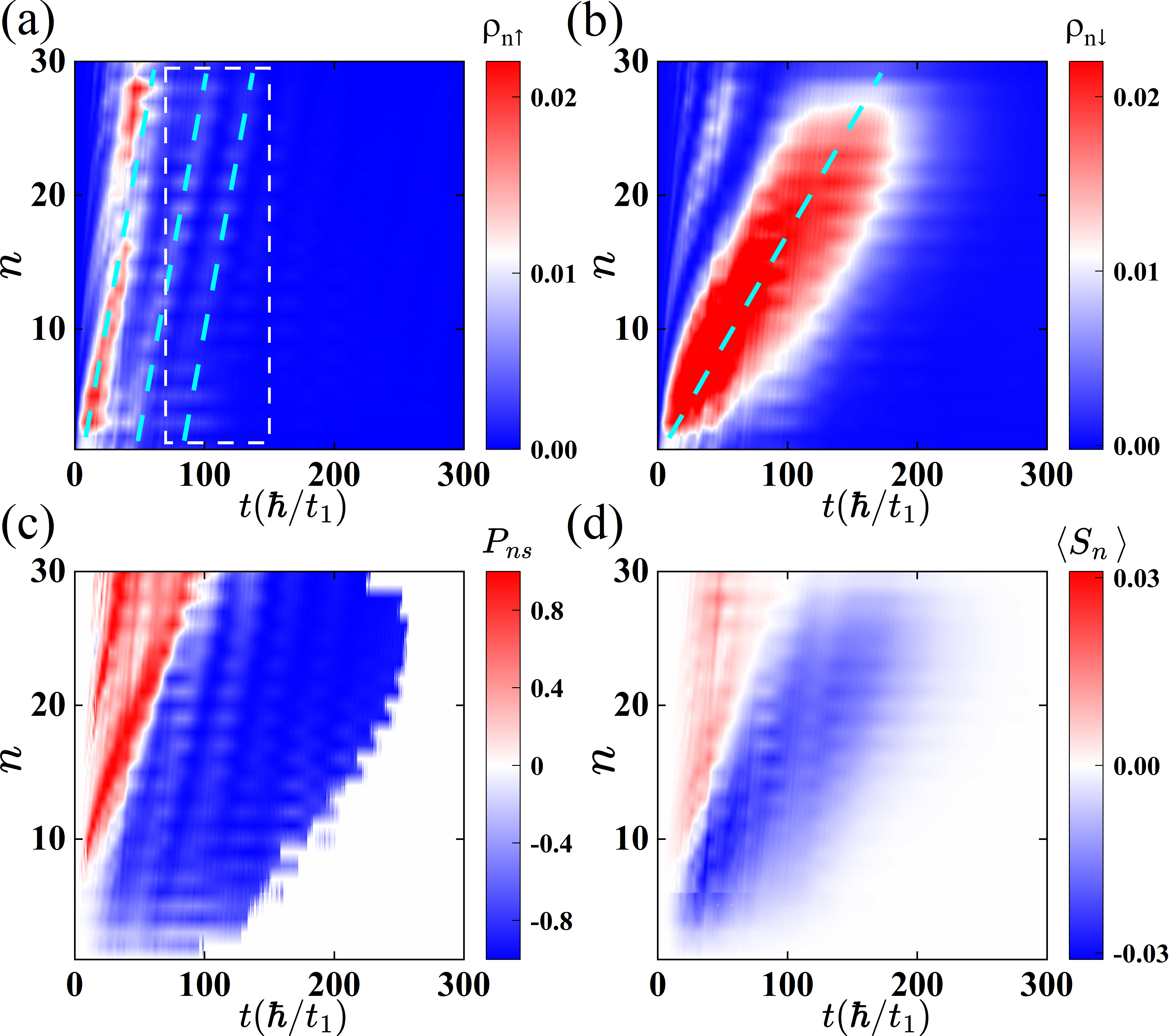}%
	\caption{\label{fig3} Time evolution of different spin and the instantaneous polarization. (a) and (b) 2D plot of spin-up density (a) and spin-down density (b) versus the time $t$ and site number $n$. Amplified figures of the regions marked by white dashed boxes in (a) is shown in Fig. \ref{fig8}(a). The trajectory of the high-density region demonstrates the velocity of spin motion. The velocity is equal to the slope of the trajectory which is illustrated by cyan-dashed lines. Electrons with spin-up have larger slopes, which indicates a higher velocity than the spin-down electrons, resulting in spin separation and instantaneous spin polarization at the endpoint A. (c) and (d) The spin polarization $P_{ns}$ (c) and average spin $\langle S_n \rangle$ (in unit of $\hbar/2$) (d) versus the site $n$ of B and the time $t$. The spin polarization $P_{ns}$ is only conducted on the sites of $\rho_{n}=\rho_{n\uparrow}+\rho_{n\downarrow}>0.001$, since it is not well-defined at the very small $\rho_{n}$. All parameters are the same as in Fig. \ref{fig2}.}
\end{figure}

\begin{figure}
	\includegraphics[width=1\columnwidth]{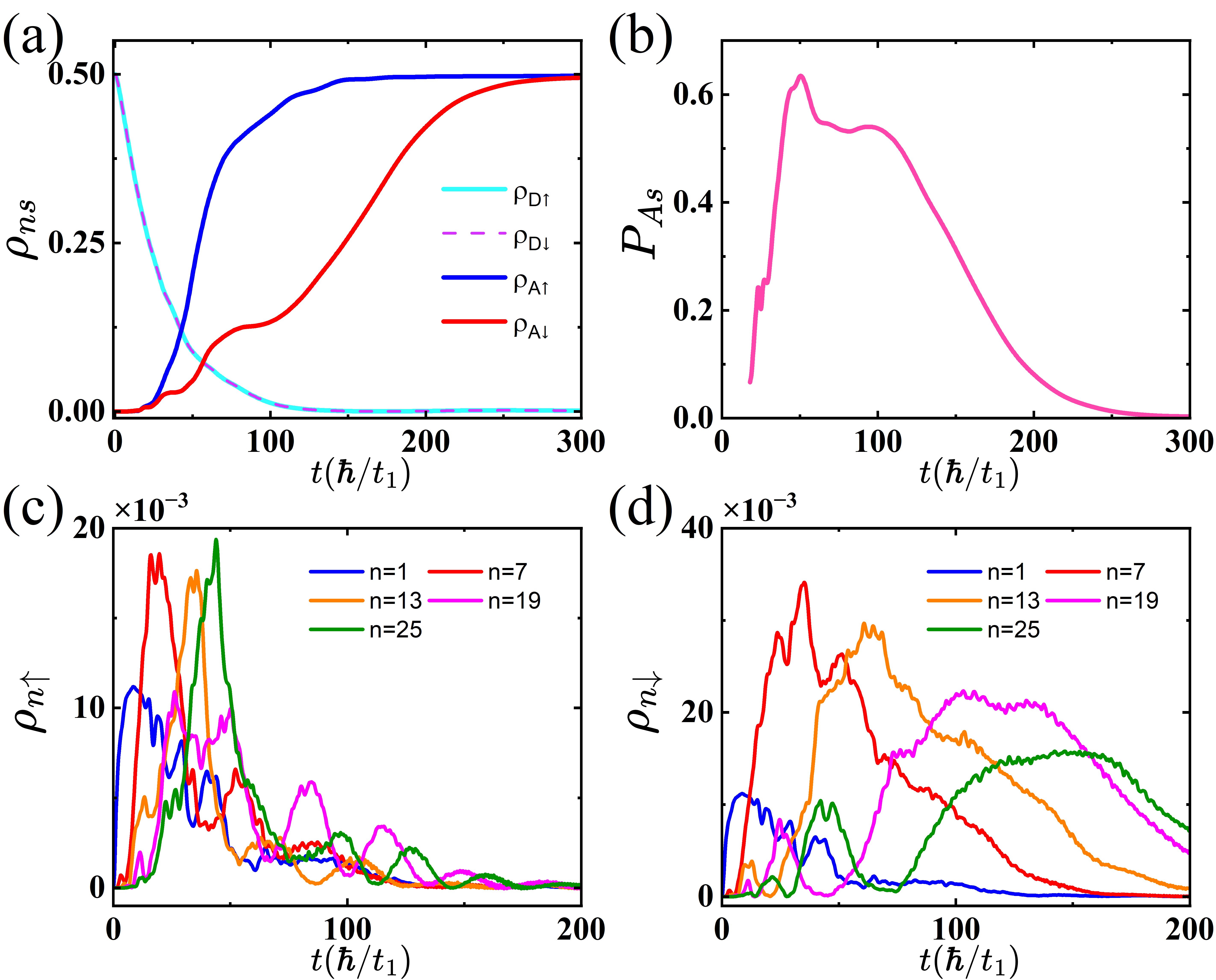}%
	\caption{\label{fig4} (a) Spin densities of D and A without dephasing. Different spins in D decay exponentially at a similar rate, but the accumulation of spin-up in A is faster, causing instantaneous polarization. (b) The spin polarization at different times in A starts at $\rho_{A}=\rho_{A\uparrow}+\rho_{A\downarrow}>0.01$. A high polarization shows up in the short time scale, but decays to zero in the long time scale. (c) and (d) The spin densities of spin-up (c) and spin-down (d) on some specific sites of B. The parameters are the same as in Fig. \ref{fig2}.}
\end{figure}

We first study the spin evolution without the dephasing process ($\Gamma_d=0$). Fig. \ref{fig2} shows the spin density $\rho_{ns}$ at each site $n$ from the time $t=0$ to $t=140\hbar/t_1$. At $t=0$, the system is in its initial state with $\rho_{D\uparrow}=\rho_{D\downarrow}=0.5$, and the polarization $P_{Ds}=0$. At $t=20\hbar/t_1$, electrons start to enter B and propagate toward A. Due to the unequal velocities, the spin-up electrons travel a greater distance, while the spin-down electrons stay closer to D due to their slower speed. At $t=40\hbar/t_1$, the majority of the spin-up electrons have arrived at position $n=25$, while the spin-down electrons remain at position $n=10$. During the period between $t=60\hbar/t_1$ and $100\hbar/t_1$, a lot of the spin-up electrons have reached the right end of B and entered A, while the majority of the spin-down electrons are still moving in the middle of B. During the period between $t=120\hbar/t_1$ and $140\hbar/t_1$, nearly all spin-up electrons have arrived at A, and the difference in the number of spin-up and spin-down electrons in A result in instantaneous high spin polarization. In the long-time range (see the Movie S1 in the Supplemental Material \cite{Sup}), the spin-down electrons slowly reach and enter A. Because the number of both spin types of electrons is equal, the spin polarization in A returns to $P_{As}=0$.

To further analyze the velocity difference of spins, Fig. \ref{fig3}(a) and Fig. \ref{fig3}(b) depict the time evolutions of $\rho_{n\uparrow}$ and $\rho_{n\uparrow}$ in B, respectively. The trajectories formed by the regions with high densities in the figures show the positions of electrons at different time $t$. Therefore, the slope of each trajectory (as illustrated by cyan-dashed lines) is the velocity. In Fig. \ref{fig3}(a), the steeper slope of the spin-up trajectory suggests a higher velocity. Conversely, the gentler slope in Fig. \ref{fig3}(b) indicates a significantly lower velocity for spin-down electrons. From Figs. \ref{fig3}(a, b), it can be observed that the overall electron transfer time is on the order of hundreds of $\hbar/t_1$, which corresponds to a few ps (with $t_1 = 0.1 \ \rm{eV}$), indicating an ultrafast spin transfer process. Local polarization $P_{ns}$ and average spin $\langle S_n \rangle = \hbar/2 (\rho_{n\uparrow}-\rho_{n\downarrow})$ for every site $n$ in B are presented in Fig. \ref{fig3}(c) and Fig. \ref{fig3}(d), respectively. Since $P_{ns}=(\rho_{n\uparrow}-\rho_{n\downarrow})/(\rho_{n\uparrow}+\rho_{n\downarrow})$ is not well-defined at the very small $\rho_{n}=\rho_{n\uparrow}+\rho_{n\downarrow}$, we start to calculate $P_{ns}$ on the sites of $\rho_{ns}>0.001$. Because the spin-up electrons move faster, they can reach positions with larger $n$ in a shorter time. Meanwhile, the slower-moving spin-down electrons are left far behind. Therefore, both $P_{ns}$ and $\langle S_n \rangle$ are positive for larger $n$ and are negative for smaller $n$ at the shorter time ($t < 100\hbar/t_1$). After some time, all the spin-up electrons pass through B, while the spin-down electrons move within B, leading to negative $P_{ns}$ and $\langle S_n \rangle$.

After analyzing the overall dynamical process, we next focus on the dynamical process of spins at specific sites. Since the spin polarizations in D and A are a key factor to yield spin-selective transfer, in Fig. \ref{fig4}(a) we show the spin densities at both sites. Because the transfer process from D to B is rapid and independent of spin, the spin densities $\rho_{D\uparrow}$ and $\rho_{D\downarrow}$ decay exponentially at a similar rate. However, electrons with different spins have different speeds when passing through B to reach A, therefore the accumulation rates of different spins in A are different. Initially, the densities of both spins in A are almost 0, and after a time of propagation, the spin-up electrons arrive at A due to their faster speed and result in a larger $\rho_{A\uparrow}$. After a while, the spin-down electrons finally arrive at A and start to accumulate, and $\rho_{A\downarrow}$ increases. It is worth noting that, although $\rho_{A\uparrow}$ and $\rho_{A\downarrow}$ are different in the short time ($t<100\hbar/t_1$), the final $\rho_{A\uparrow}$ and $\rho_{A\downarrow}$ are the same at the long time and become unpolarized. Fig. \ref{fig4}(b) shows the spin polarization $P_{As}$ of A. Since $P_{As}=(\rho_{A\uparrow}-\rho_{A\downarrow})/(\rho_{A\uparrow}+\rho_{A\downarrow})$ is not well-defined at the very small $\rho_A=\rho_{A\uparrow}+\rho_{A\downarrow}$, we start to calculate $P_{As}$ from $\rho_A>0.01$. At the arrival of electrons, the spin polarization $P_{As}$ first rapidly rises to over 60\%, then remains highly polarized for some time, and then decays to 0 at the arrival of spin-down electrons. In addition, the distributions of $\rho_{n\uparrow}$ and $\rho_{n\downarrow}$ on specific sites in B are shown in Fig. \ref{fig4}(c) and Fig. \ref{fig4}(d), respectively. From these figures, it can be observed that the spin-up and spin-down electrons have higher and slower velocities, respectively.

The instantaneous high spin polarization may capture the essence of the ultrafast charge transfer process in the radical pair separation experiments: the polarization arises from the unequal spin velocities because of the SOC and helical structure and has nothing to do with the substrates or electrodes \cite{Eckvahl2023}.

\begin{figure}
	\includegraphics[width=1\columnwidth]{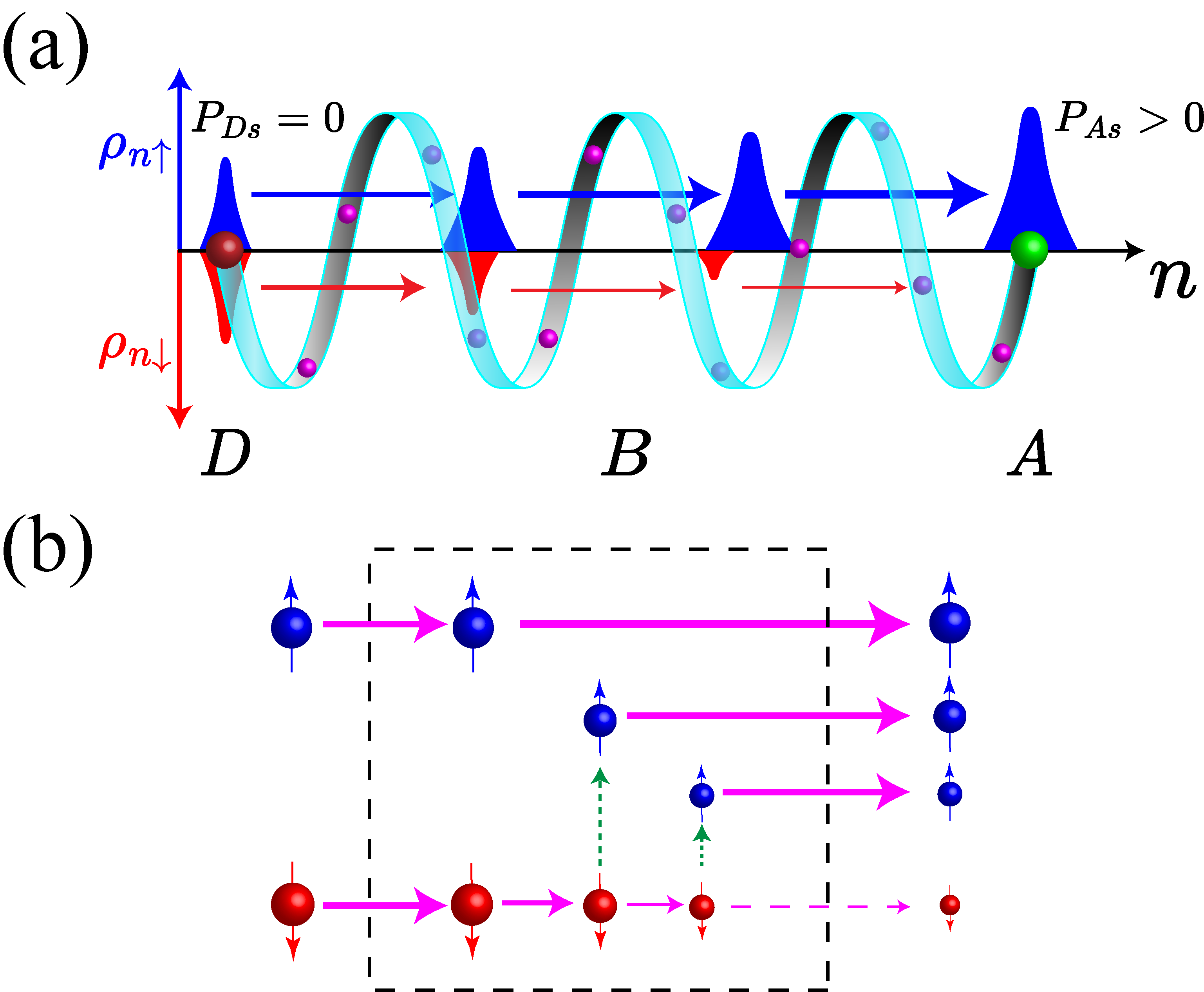}%
	\caption{\label{fig5} (a) Schematics of the spin propagation and conversion. Spins with equal densities start at D, and the unequal velocities induce a spin separation and a conversion from the slow spin (spin-down) to the fast spin (spin-up). This spin conversion leads to unequal spin densities and steady spin polarization. (b) Schematics of the spin conversion mechanism. The same number of opposite spins are traveling through the black dashed box (chiral molecules) area. Because it takes less time for the fast spin (blue) to pass the area, they will suffer less dephasing. The slow spin (red) will stay in the area and suffer dephasing events, then convert to the fast spin. Once they turn into a fast spin, they will leave the region with high velocity.}
\end{figure}

\begin{figure}
	\includegraphics[width=1\columnwidth]{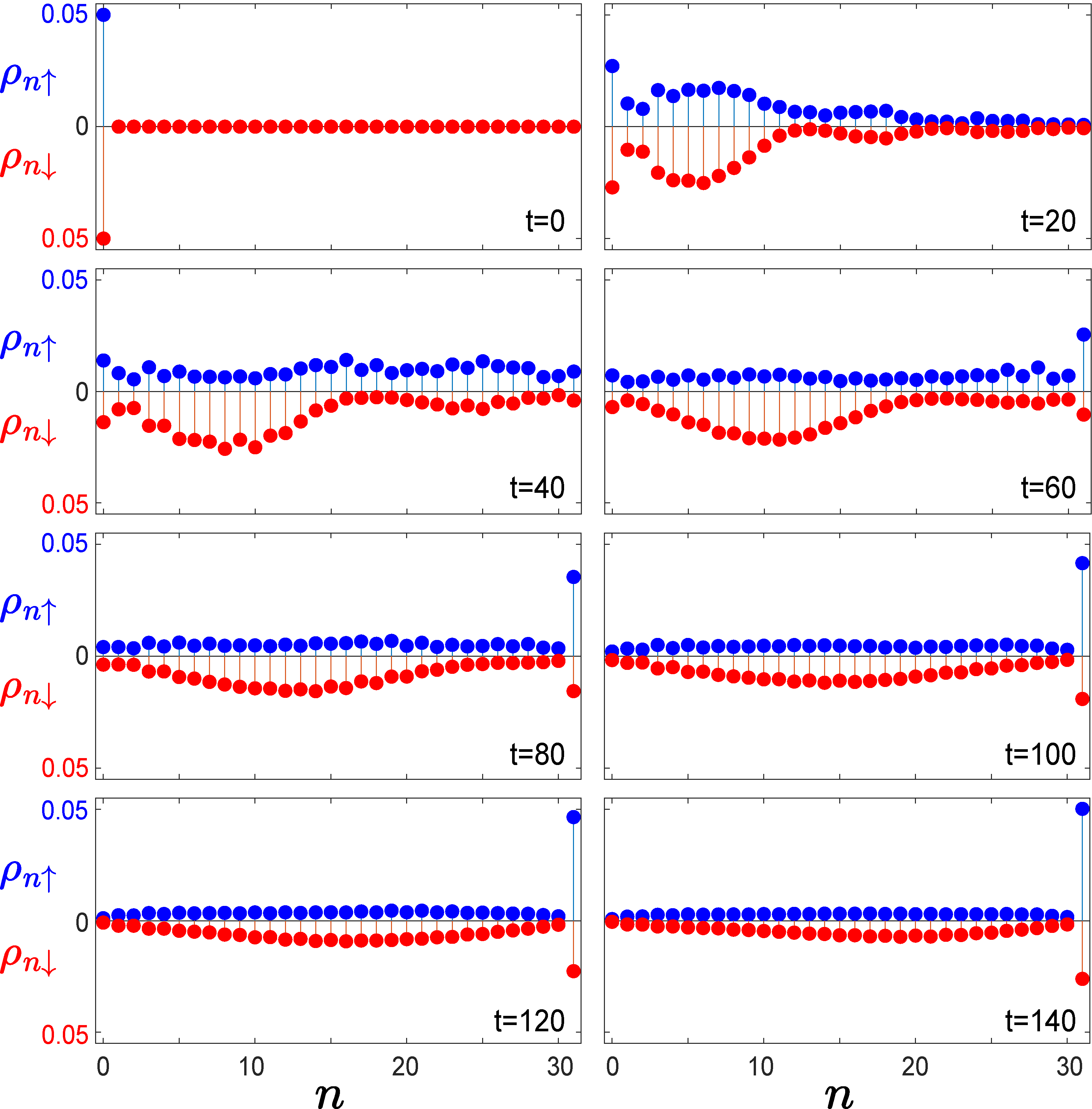}%
	\caption{\label{fig6} Dynamical process of spin transfer with dephasing. The spin density at D ($n=0$) and A ($n=N+1=31$) has been reduced to one-tenth of its original value for better illustration, and the time labeled in the figures is in unit of $\hbar/t_1$. All parameters are the same as Fig. \ref{fig2} except $\Gamma_d=0.005t_1$.}
\end{figure}

\begin{figure}
	\includegraphics[width=1\columnwidth]{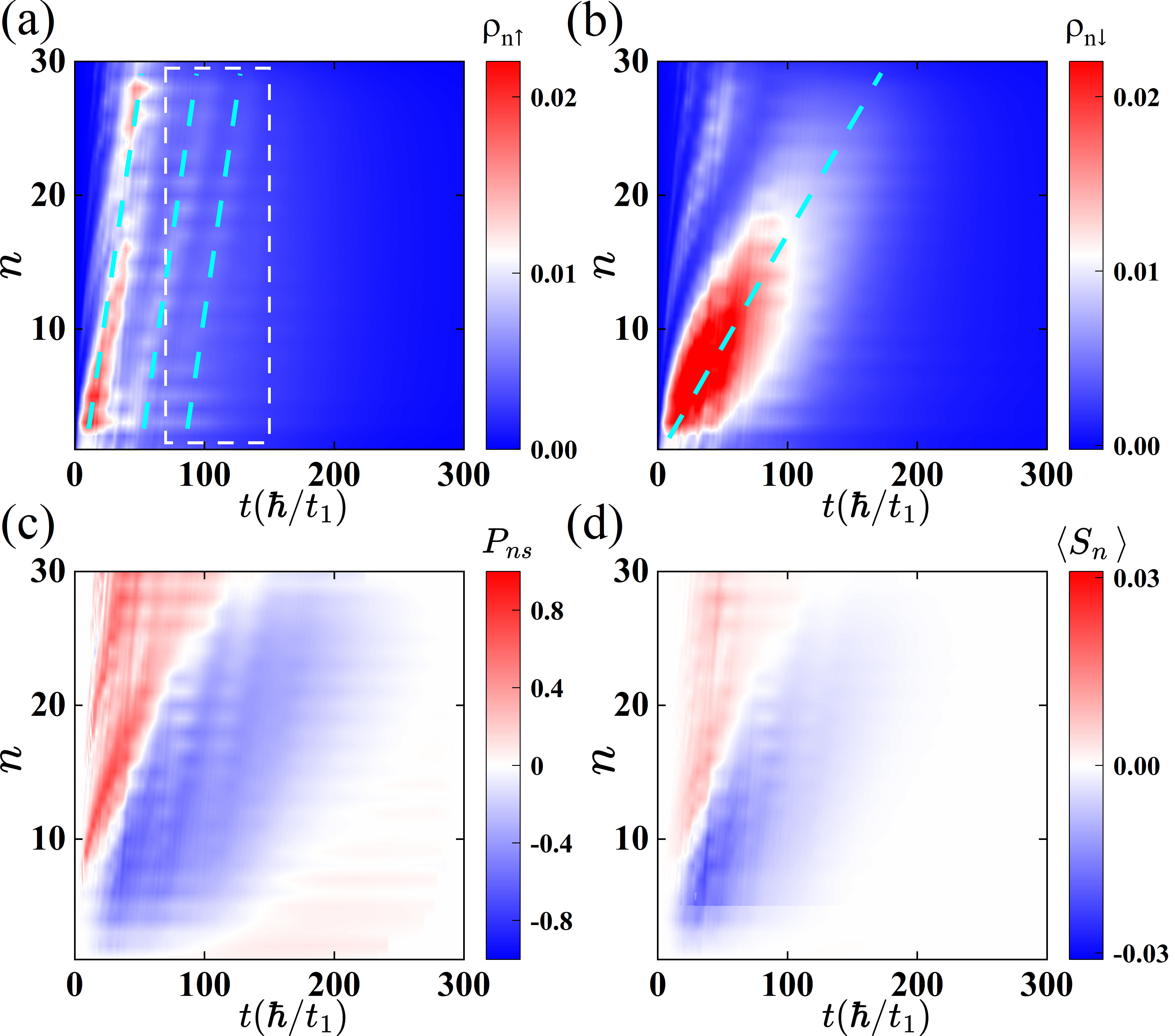}%
	\caption{\label{fig7} The spin conversion induced by the dephasing process.
(a) and (b) 2D plot of spin-up density (a) and spin-down density (b) versus the time $t$ and site number $n$ with the dephasing process. Amplified figures of the regions marked by white dashed boxes in (a) is shown in Fig. \ref{fig8}(b). The spin velocities are equal to the slope of the trajectory, which are illustrated by cyan-dash lines. The increase in spin-up density and the decrease in spin-down density (compared to Fig. \ref{fig3}(a) and Fig. \ref{fig3}(b)) over a long period imply spin conversion caused by dephasing. (c) and (d) The spin polarization $P_{ns}$ (c) and average spin $\langle S_n \rangle$ (in unit of $\hbar/2$) (d) versus the site $n$ of B and the time $t$. The spin polarization $P_{ns}$ is only conducted on the sites of $\rho_{n}=\rho_{n\uparrow}+\rho_{n\downarrow}>0.001$, since it is not well-defined at the very small $\rho_{n}$. All parameters are the same as in Fig. \ref{fig6}.}
\end{figure}

\begin{figure}
	\includegraphics[width=1\columnwidth]{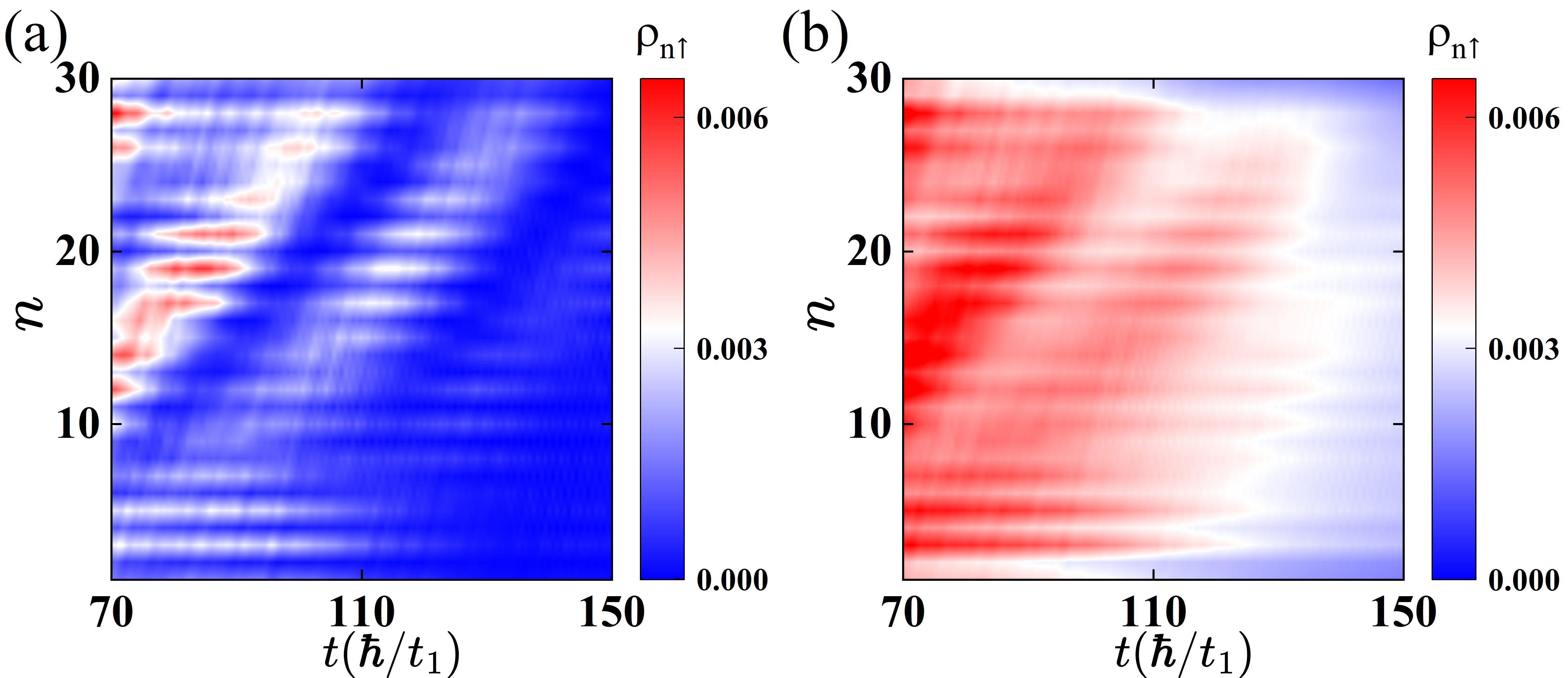}%
	\caption{\label{fig8} Comparison of spin-up evolution without ($\Gamma_d=0$) and with dephasing ($\Gamma_d=0.005t_1$). (a) Locally amplified figure of Fig. \ref{fig3}(a) during $t=70 \sim 150 \hbar/t_1$. The majority of the up-spin electrons have moved to region A, so the density of up-spin electrons in this area is low. (b) Locally amplified figure of Fig. \ref{fig7}(a) during $t=70 \sim 150 \hbar/t_1$. Due to the dephasing process, spin-down electrons are continuously converted into spin-up electrons, causing significantly large densities of spin-up electrons during the same period.}
\end{figure}

\section{\label{SEC5} Dephasing induced steady spin polarization}

We then study the influence of the dephasing process on the spin transport along the chiral molecule by setting $\Gamma_d = 0.005 t_1$ and show that this could lead to long-term steady spin polarization. This process is described by the third term on the right-hand side of Eq. (\ref{ME}), which causes electrons to lose their phases and spin memories. In the process of the spin memory loss, the average electron spin $\langle S_n \rangle$ at each position $n$ decays exponentially to 0 with lifetime $\hbar/\Gamma_d$. Consequently, if $\rho_{n\uparrow}>\rho_{n\downarrow}$, dephasing will induce the conversion of spin-up electrons to spin-down ones. Conversely, if $\rho_{n\uparrow}<\rho_{n\downarrow}$, spin-down electrons will be converted to spin-up ones.

The physical picture of the electron transfer process under the influence of the dephasing process is shown in Fig. \ref{fig5}(a) and Fig. \ref{fig5}(b). It will be shown that dephasing is the driving force of a spin-converting phenomenon in the dynamical process \cite{Guo2014}. Consider a same number of spin-up and spin-down electrons travel through the area marked by a black dashed box (representing chiral molecules) in Fig. \ref{fig5}(b). The spin-up electrons with high velocities experience less dephasing and will maintain their spin direction as they quickly pass through this area. Consequently, this area is full of slow spin-down electrons and the average electron spin is negative, $\langle S_n \rangle < 0$. Under the influence of dephasing, $\langle S_n \rangle$ will exponentially decay to 0, indicating that spin-down electrons convert to spin-up electrons. Once they are converted into spin-up electrons, they will quickly leave this area; conversely, if the spin is preserved, they will remain in the area and experience further dephasing [see Fig. \ref{fig5}(b)]. This spin conversion process causes spin-up electrons to be more than spin-down electrons and results in a steady spin polarization shown in Fig. \ref{fig5}(a).

Next, we numerically study the influence of dephasing on the spin dynamical process of electrons transferring through the chiral molecule, as shown in Fig. \ref{fig6}. During the time between $t=0$ and $t=20\hbar/t_1$, the same number of spin-up and spin-down electrons start to transfer from D to B. Because the time of motion is short, the dephasing effect has not fully manifested, so the motion pattern is similar to the one in Fig. \ref{fig2} during $t=0 \sim 20 \hbar/t_1$. At time $t=40 \hbar/t_1$, because the spin-up electrons move faster, they have reached $n=25$, while the majority of spin-down electrons have just moved to $n=10$. Compared to the same time in Fig. \ref{fig2} (no dephasing), $\rho_{n\uparrow}$ in Fig. \ref{fig6} at positions $n<15$ is larger than that in Fig. \ref{fig2}, while $\rho_{n\downarrow}$ is smaller than that in Fig. \ref{fig2}. This indicates that the spin-down electrons in Fig. \ref{fig6} have been converted into spin-up electrons due to the dephasing process, consistent with the physical picture presented above. During time $t=60 \sim 100 \hbar/t_1$, the spin separation and conversion occur simultaneously. A lot of spin-up electrons have reached A, but the majority of spin-down electrons remain at $n=10 \sim 20$. Compared to the same period in Fig. \ref{fig2}, the spin-down electrons in Fig. \ref{fig6} are continuously converting into spin-up electrons, leading to a decrease in the density of spin-down electrons. Finally, during time $t=120 \sim 140 \hbar/t_1$, the majority of spin-up electrons have arrived at A, but the density of spin-down electrons in A is still small. Because of the spin conversion, the total number of spin-up electrons is larger than that of spin-down electrons. Therefore, even if all spin-down electrons reach A in a long time range (see the Movie S2 in the Supplemental Material \cite{Sup}), steady spin polarization can still be generated in A.

The time evolutions of $\rho_{n\uparrow}$ and $\rho_{n\downarrow}$ in B and the corresponding polarization $P_{ns}$ and average spin $\langle S_n \rangle$ under the influence of dephasing are shown in Figs. \ref{fig7}(a-d). The cyan-dashed lines in Figs. \ref{fig7}(a, b) illustrate the slope of the trajectories, which are velocities of spin-up and spin-down electrons. From these two figures, it can be observed that the slope lines in Fig. \ref{fig7}(a) (spin-up electrons) are steeper than that in Fig. \ref{fig7}(b) (spin-down electrons). This indicates that spin-up electrons possess a higher velocity, which is consistent with the results shown in Figs. \ref{fig3}(a, b). Therefore, it can be concluded that dephasing does not alter the velocity of electrons at all; its influence on electron dynamics is exerted by changing the population of electrons with different velocities through spin conversion. The electron conversion process can be observed by comparing Figs. \ref{fig7}(a, b) with Figs. \ref{fig3}(a, b). Although the trajectory shapes in Fig. \ref{fig7}(a) and Fig. \ref{fig3}(a) are similar, their densities exhibit significant differences, especially at longer motion times ($t>70\hbar/t_1$), where the density in Fig. \ref{fig7}(a) is notably higher than that in Fig. \ref{fig3}(a). For better comparison, the amplified figures of the regions marked by white dashed boxes in Fig. \ref{fig3}(a) and Fig. \ref{fig7}(a) are shown in Fig. \ref{fig8}(a) and Fig. \ref{fig8}(b), respectively. It is clear that $\rho_{n\uparrow}$ during time $t= 70 \sim 150 \hbar/t_1$ in Fig. \ref{fig7}(a) [Fig. \ref{fig8}(b)] is significantly higher than that in Fig. \ref{fig3}(b) [Fig. \ref{fig8}(a)]. This significant increase in the population of spin-up electrons in Fig. \ref{fig8}(b) is attributed to the conversion of spin-down electrons to spin-up electrons induced by dephasing. In Fig. \ref{fig7}(b), the spin-down density becomes lower and lower as time passes, as it consistently flips into the fast spin-up state. The influence of dephasing is also evident in Figs. \ref{fig7}(c, d). Their trajectory shapes are similar to those in Figs. \ref{fig3}(c, d) ($\Gamma_d = 0$), but both $P_{ns}$ and $\langle S_n \rangle$ are lower. This is because both spin states are affected by dephasing, which causes $\rho_{n\uparrow} - \rho_{n\downarrow}$ to decay.

\begin{figure}
	\includegraphics[width=1\columnwidth]{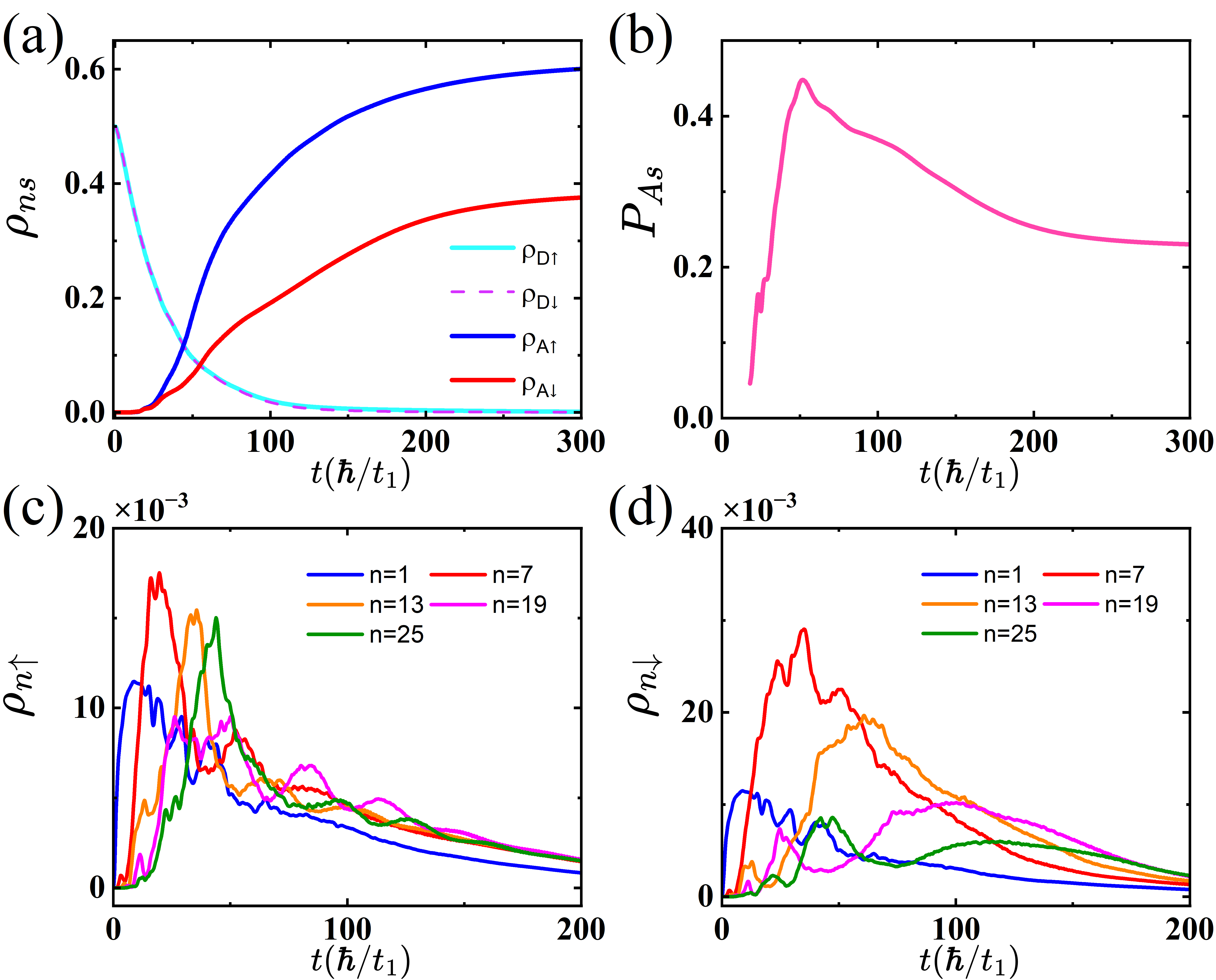}%
	\caption{\label{fig9} (a) Spin densities of D and A with dephasing. The dephasing process causes a great difference between spin-up and spin-down densities in A, resulting in a steady spin polarization. (b) Spin polarization in A with dephasing, starts at $\rho_{A}=\rho_{A\uparrow}+\rho_{A\downarrow}>0.01$. The spin separation leads to instantaneous high polarization, and the spin conversion leads to long steady spin polarization. (c) and (d) The spin densities of spin-up (c) and spin-down (d) on some specific sites of B with dephasing.
All parameters are the same as in Fig. \ref{fig6}.}
\end{figure}

The dynamics of spin evolution at specific sites under dephasing is illustrated in Fig. \ref{fig9}. Fig. \ref{fig9}(a) depicts the evolution of spin density in D and A. Although the two spins in D undergo the same exponential decay as in Fig. \ref{fig4}(a), the spin accumulation in A are completely different. The spin conversion induced by dephasing results in a greater number of spin-up electrons, leading to an unequal $\rho_{A\uparrow}$ and $\rho_{A\downarrow}$ at longer times ($t>250 \hbar/t_1$), with $\rho_{A\uparrow}$ exceeding 0.5 while $\rho_{A\downarrow}$ remains below 0.5. As time progresses further, more electrons enter A, yet the difference in spin density remains; this results in a long-term stable spin polarization in A. To confirm this, in Fig. \ref{fig9}(b) we show the spin polarization $P_{As}$ versus the time $t$. $P_{As}$ initially rises rapidly to a high value, reaching a peak before starting to decline. During the shorter time ($t=20 \sim 50 \hbar/t_1$), primarily spin-up electrons enter A, while spin-down electrons remain in B, and this behavior is similar to that in Fig. \ref{fig4}(b). However, in the long time period, $P_{As}$ does not decay to zero as in Fig. \ref{fig4}(b) but instead stabilizes at over 20\%; this stabilization is driven by spin conversion. Since the balance of spin populations is disrupted by dephasing, the total number of electrons with different spins in the system no longer remains equal, leading to a stable polarization in A.
Meanwhile, Fig. \ref{fig9}(c) and Fig. \ref{fig9}(d) show the distributions of $\rho_{n\uparrow}$ and $\rho_{n\downarrow}$ on specific sites, respectively. At longer times ($t > 70\hbar/t_1$), the spin-up electron density in Fig. \ref{fig9}(c) shows an increase compared to that in Fig. \ref{fig4}(c), while the spin-down electron density in Fig. \ref{fig9}(d) exhibits a decrease compared to Fig. \ref{fig4}(d). This further demonstrates the conversion of spin-down electrons into spin-up electrons from another perspective.

\begin{figure}
	\includegraphics[width=1\columnwidth]{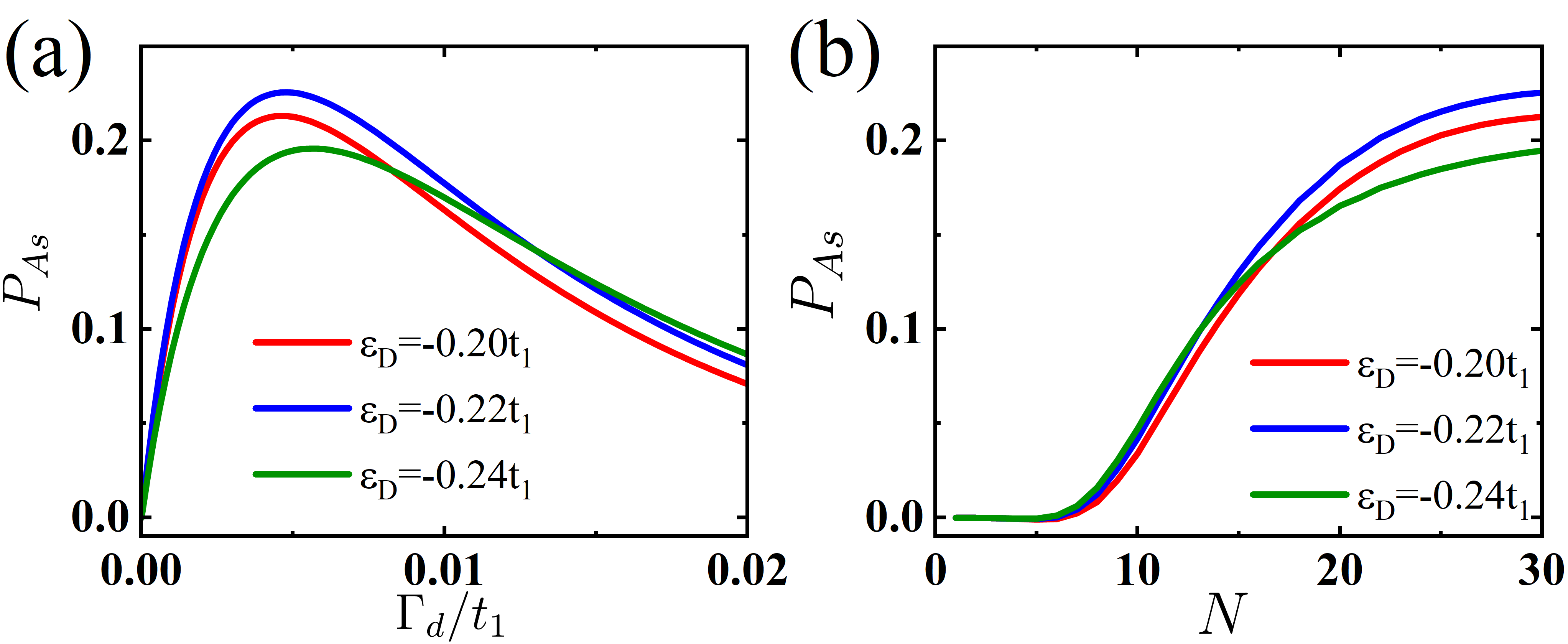}%
	\caption{\label{fig10} (a) Steady polarization $P_{As}$ versus dephasing strength $\Gamma_d$ with $N=30$. The steady spin polarization is chosen at time $t=1000\hbar/t_1$. (b) $P_{As}$ versus molecular length $N$ with $\Gamma_d=0.005 t_1$. Other parameters are the same as those in Fig. \ref{fig6}. The spin polarization effect is significant in a wide range of model parameters.}
\end{figure}

Figure \ref{fig10}(a) and Fig. \ref{fig10}(b) show the steady spin polarization $P_{As}$ versus the dephasing strength $\Gamma_d$ and molecular length $N$ in different energy levels $\varepsilon_D$ of D, respectively. The steady spin polarization $P_{As}$ is evaluated at time $t=1000 \hbar/t_1$ ($\approx$ 6.6 ps) as this time scale is much longer than the electron transfer time. As shown in Fig. \ref{fig10}(a), when the dephasing strength is relatively small, the steady spin polarization is nearly zero because of the nearly equal number of both spins. With the increase in dephasing strength, the spin conversion becomes significant, resulting in considerable polarization. However, as the dephasing strength continues to increase, the steady spin polarization begins to decline, because the strong dephasing completely breaks the spin memory of both spins. The maximum steady spin polarization occurs at $\Gamma_d \approx 0.005 t_1$, as the average spin dephasing lifetime, $\hbar/\Gamma_d \approx 200 \hbar/t_1$, falls precisely between the spin-up and spin-down electron transfer times. Consequently, spin-up electrons remain largely unaffected by dephasing, whereas spin-down electrons experience significant dephasing effects. Fig. \ref{fig10}(b) shows that with the increase of molecular length $N$, the steady spin polarization continues to increase. This is because the longer the propagation time, the larger the spin separation will be. The dependences of the polarization on $\Gamma_d$ and $N$ are well consistent with previous theories and experiments \cite{Guo2012_2,Mishra2020}, indicating a high degree of universality in the dynamical theory here.

\section{\label{SEC6} Discussions and conclusions}

In the paper, the electron velocity is qualitatively described through the slope of trajectories formed by regions with high electron densities in Fig. \ref{fig3} and Fig. \ref{fig7} (as indicated by cyan-dashed lines). 
For non-helical chiral molecules, the arrangement of lattice points is irregular, 
but the evolution of the electron density can still reflect the electron ``velocity". For these molecules, electrons propagating from D to the A still sequentially pass through intermediate units (atoms, nucleobases, or amino acids) in the chiral molecular bridge during their transport. 
Although the path does not follow a linear or helical trajectory, the electrons still propagate along this sequential pathway composed of intermediate units, and the densities' site number versus time reflects the electron transport. This transport process of electron spin from D to A can still be qualitatively characterized and distinguished using velocity-related descriptions. 

Besides, the protein-like molecule is a special type of chiral molecules exhibiting a vanishing HOMO-LUMO gap as the chain gets longer--a consequence of its single-helical structure, which generate a single electronic band with the Fermi level lying inside it \cite{Guo2014,Guo2014}. However, many molecules have a fixed HOMO-LUMO gap; to rigorously demonstrate the universality of our dynamical theory, we further extend our analysis to a distinct chiral system: dsDNA, which maintains a fixed HOMO-LUMO gap \cite{Guo2012,Guo2012_2,Guo2014_1}. The Hamiltonian and dynamical equation of the D-dsDNA-A systems are shown in the Appendix, and the computational results are summarized in Fig. \ref{fig11}, which behavior are similar to protein-like molecules. The agreement across these divergent chiral systems validates the robustness and general applicability of our theoretical framework.

In summary, we have developed a spin dynamical theory of the CISS effect using a Lindblad-type master equation and a clear physical picture of unequal spin velocities induced by molecular SOC and chiral structure. This method was applied to a typical D-B-A system without electrodes or substrates. In the absence of the dephasing process, instantaneous high spin polarization is found in A and vanishes in the long time. While in the presence of the dephasing process, a steady spin polarization shows up because of the combination of unequal velocities and spin conversion. Our results fill in the gaps in the CISS theory and may facilitate the engineering of chiral-based spintronic devices.

\begin{acknowledgments}
    This work was financially supported by
    the National Key R and D Program of China (Grant No. 2024YFA1409002),
    the National Natural Science Foundation of China (Grant No. 12374034, No. 11921005, No. 12274466),
    the Innovation Program for Quantum Science and Technology (2021ZD0302403), and the Hunan Provincial Science Fund for Distinguished Young Scholars (Grant No. 2023JJ10058). We also acknowledge the High-performance Computing Platform of Peking University for providing computational resources.
\end{acknowledgments}

\appendix
\section*{Appendix: Dynamical process of double-stranded DNA (dsDNA)}

In this Appendix, we take the dsDNA as the chiral molecule bridge B 
in the D-B-A system to validate the robustness and general applicability of our theoretical framework.
The D-dsDNA-A system can be simulated 
by the Hamiltonian 
$H_{DNA} = H_B^{DNA} +H_{BD}^{DNA} +H_A^{DNA}$. 
The first term $H_B^{DNA}$ is the Hamiltonian of the dsDNA bridge \cite{Guo2012}:
\begin{widetext}
\begin{eqnarray}
    H_B^{DNA}&=&\left[
    \sum_{n=1}^{N_{DNA}} \sum_{j=1}^{2}\varepsilon_{jn}^{DNA}c_{jn}^{\dagger}c_{jn}
    +\sum_{n=1}^{N_{DNA}-1}\sum_{j=1}^{2}t_{jn}^{DNA}c_{jn}^{\dagger}c_{jn+1}+\sum_{n=1}^{N_{DNA}}\lambda_{n}^{DNA}c_{1n}^{\dagger}c_{2n}+h.c.\right] 
    \nonumber\\
    &+&\sum_{n=1}^{N_{DNA}-1}\sum_{j=1}^{2}\left(it_{so}^{DNA}c_{jn}^{\dagger}\left[\sigma_{n}^{(j)}+\sigma_{n+1}^{(j)}\right]c_{jn+1}\right),
     \label{HB_DNA}
\end{eqnarray}
\end{widetext}
where $c_{jn}^{\dagger}=(c_{jn\uparrow}^{\dagger},c_{jn\downarrow}^{\dagger})$ is the creation operator of the electron at the $n$th site of the $j$th chain of the dsDNA. The DNA length is $N_{DNA}$. 
$\varepsilon_{jn}^{DNA}$, $t_{jn}^{DNA}$, $\lambda_{n}^{DNA}$, and $t_{so}^{DNA}$
are the on-site energy, the intrachain hopping integral, interchain hopping integral, and the SOC, respectively. 
Defining $\sigma_{\perp}(\varphi_{DNA})=\sigma_x \sin \varphi_{DNA} \sin \theta_{DNA}-\sigma_y \cos \varphi_{DNA} \sin \theta_{DNA}+\sigma_z \cos \theta_{DNA}$, here $\theta_{DNA}$ and $\varphi_{DNA}$ are the helix angle and the cylindrical coordinate, respectively, then $\sigma_{n+1}^{(1)}=\sigma_{\perp}(n\Delta \varphi_{DNA})$ and $\sigma_{n+1}^{(2)}=\sigma_{\perp}(n\Delta \varphi_{DNA}+\pi)$. $\Delta \varphi_{DNA}$ is the twist angle between successive base-pairs. 
The structure parameters of the dsDNA are pitch $h_{DNA}$, radius $R_{DNA}$, and arc length $l_{DNA}$, and they satisfy $l_{DNA} \cos \theta_{DNA}=R_{DNA} \Delta \varphi_{DNA}$ and $l_{DNA} \sin \theta_{DNA}= \Delta h_{DNA}$, with $\Delta h_{DNA}$ the stacking distance between neighboring base-pairs. 
The second term $H_{BD}^{DNA}=\varepsilon_{D}^{DNA}c_{D}^{\dagger}c_{D}+\sum_{j=1}^{2}(t_{BD}^{DNA}c_{j1}^{\dagger}c_{D}+h.c.)$ in $H_{DNA}$ is the Hamiltonian of donor D and the hopping between D and dsDNA, where $c_{D}^{\dagger}=(c_{D\uparrow}^{\dagger}, c_{D\downarrow}^{\dagger})$ is the creation operator of D. $\varepsilon_{D}^{DNA}$ is the on-site energy of D and $t_{BD}^{DNA}$ is the hopping integral between D and dsDNA. The last term $H_{A}^{DNA}=\varepsilon_{A}^{DNA} c_{A}^{\dagger}c_{A}$ in $H_{DNA}$ is the Hamiltonian of acceptor A, where $c_{A}^{\dagger}=(c_{A\uparrow}^{\dagger}, c_{A\downarrow}^{\dagger})$ is the creation
operator of A, and $\varepsilon_{A}^{DNA}$ is the on-site energy of A.

\begin{figure}
	\includegraphics[width=1\columnwidth]{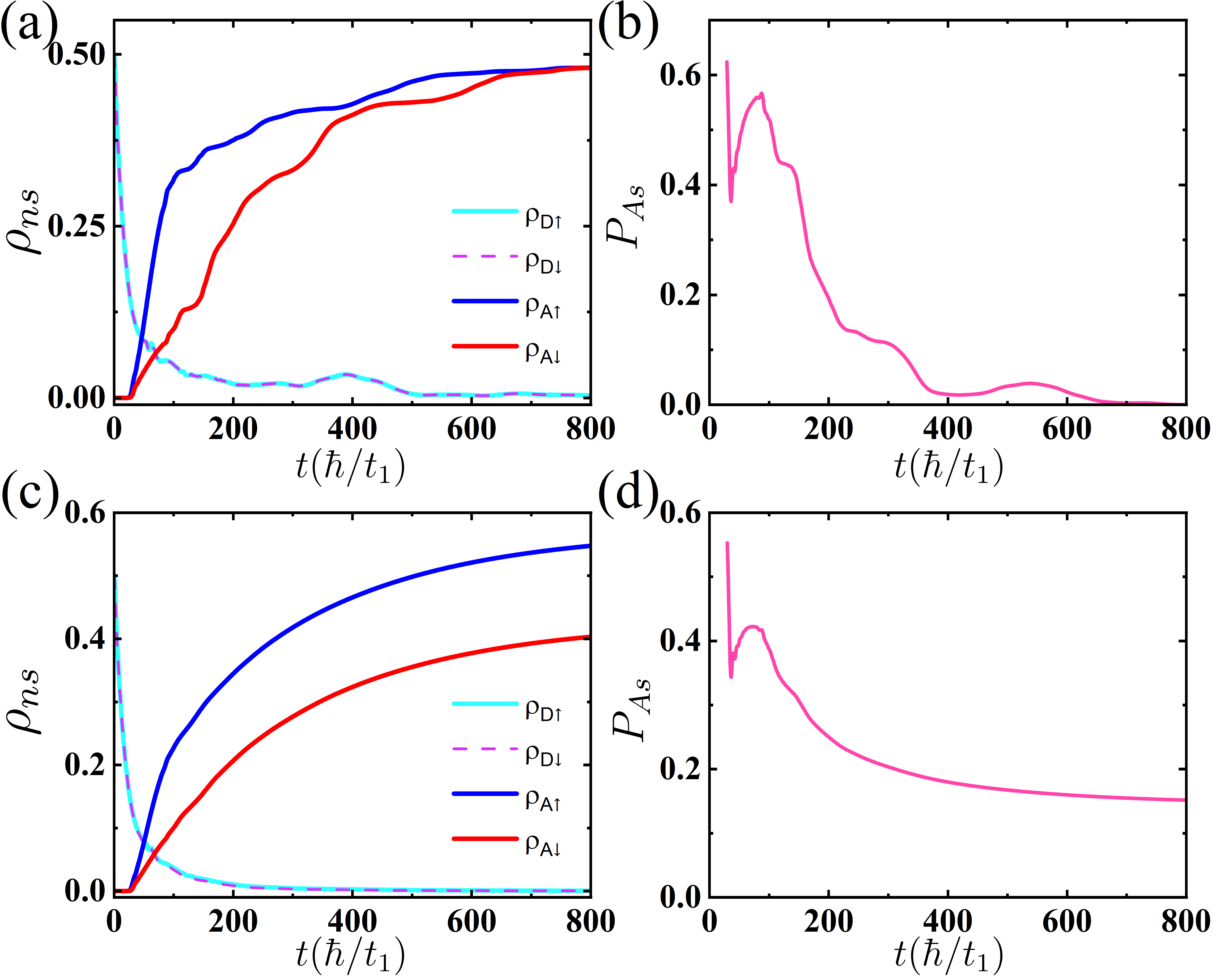}
	\caption{\label{fig11} The dynamical process of D-dsDNA-A systems. (a,b) Spin densities of D and A (a) and spin polarization in A (b) versus the time $t$ without dephasing ($\Gamma_d^{DNA}=0$). 
    (c,d) Spin densities of D and A (c) and spin polarization in A (d) versus the time $t$ with dephasing ($\Gamma_d^{DNA}=0.005t_1$). 
    The spin polarization in (b,d) is only conducted at $\rho_{A}=\rho_{A\uparrow}+\rho_{A\downarrow}>0.01$. 
    The length of dsDNA is $N_{DNA}  = 30$, the SOC is $t_{so}^{DNA}=0.03t_1$. Other parameters are $\varepsilon_{D}^{DNA}=4.68t_1$, $\varepsilon_{A}^{DNA}=-t_1$, $t_{BD}^{DNA}=0.4t_1$ and $\Gamma^{DNA}=10t_1$.}
\end{figure}

The dynamical process of spin transfer through the D-dsDNA-A system can be described by the Lindblad-type master equation \cite{Macaluso2023,Haberkorn1976,Breuer2007}:
\begin{widetext}
\begin{equation}
    \begin{aligned}
        \hbar \frac{d \rho}{d t}=&-i [H_{DNA},\rho]+\Gamma^{DNA} \left(L_{AB}^{DNA}\rho L_{AB}^{DNA \dagger} -\frac{1}{2} \{L_{AB}^{DNA \dagger} L_{AB}^{DNA}, \rho\}\right)\\
        &+\Gamma_{d}^{DNA} \sum_{n=1}^{N_{DNA}} \left[\sum_{j=1}^{2} \sum_{\mu = x,y,z}\left(L_{jn\mu} \rho L_{jn\mu}^\dagger -\frac{1}{2} \{L_{jn\mu}^\dagger L_{jn\mu}, \rho\}\right)\right],
        \label{ME2}
    \end{aligned}
\end{equation}
\end{widetext}
where $\rho$ and $H_{DNA}$ are the density matrix and the Hamiltonian of the D-dsDNA-A system. Here, $L_{AB}^{DNA}=c_{A}^{\dagger}(c_{1N_{DNA}}+c_{2N_{DNA}})$, $L_{jn\mu}=c_{jn}^{\dagger}\sigma_{\mu}c_{jn}$. $\Gamma^{DNA}$ and $\Gamma_{d}^{DNA}$ are jumping strength from the right end of B to A and dephasing strength in B, respectively. The electrons are initially localized in D and in a mixed state $\rho(0)=(|D \uparrow \rangle \langle D \uparrow |+|D \downarrow \rangle \langle D \downarrow |)/2$, which is spin unpolarized.

For the D-dsDNA-A system, the structural parameters for dsDNA are the radius $R_{DNA}  = 0.7 \ \rm{nm}$, twist angle $\Delta \varphi_{DNA}=\pi/5$, and the stacking distance $\Delta h = 0.34 \ \rm{nm}$ \cite{Guo2012}. The length of the dsDNA is $N_{DNA}  = 30$, the arc length $l_{DNA} \approx 0.56 \ \rm{nm}$, and helix angle $\theta_{DNA} \approx 0.66 \ \rm{rad}$ \cite{Guo2012}. $\varepsilon_{jn}^{DNA}$ is set to $\varepsilon_{1n}^{DNA}=0$ and $\varepsilon_{2n}^{DNA}=3t_1$, $t_{jn}^{DNA}$ is taken as $t_{1n}^{DNA}=1.2t_1$, $t_{2n}^{DNA}=-t_1$, and $\lambda_{n}^{DNA}=-3t_1$ \cite{Guo2012}. 
The SOC is $t_{so}^{DNA}=0.03t_1$. The on-site energy of D is set as $\varepsilon_{D}^{DNA}=4.68t_1$ and the hopping between D and dsDNA is $t_{BD}^{DNA}=0.4t_1$. The on-site energy of A is set as $\varepsilon_{A}^{DNA}=-t_1$ and the jumping strength is $\Gamma^{DNA}=10t_1$.

Figure \ref{fig11}(a,b) shows the case without dephasing ($\Gamma_d^{DNA}=0$). In Fig. \ref{fig11}(a), spin densities ($\rho_{D\uparrow}$ and $\rho_{D\downarrow}$) in D decay at same rates. However, acceptor A exhibits different spin accumulation densities ($\rho_{A\uparrow}$ and $\rho_{A\downarrow}$), yielding instantaneous high spin polarization in A. In the long time, $\rho_{A\uparrow}$ and $\rho_{A\downarrow}$ equilibrate, resulting in complete spin unpolarized. Fig. \ref{fig11}(b) shows the spin polarization $P_{As}$ of A. $P_{As}$ shows a high value in the short time, and then decays to 0 in the long time. 

Figure \ref{fig11}(c,d) shows the case with dephasing ($\Gamma_d^{DNA}=0.005t_1$). In Fig. \ref{fig11}(c), although two spins in D undergo the same decay as in Fig. \ref{fig11}(a), the spin conversion induced by dephasing results in a greater number of spin-up electrons, leading to an unequal $\rho_{A\uparrow}$ and $\rho_{A\downarrow}$ at longer times, with $\rho_{A\uparrow}$ exceeding 0.5 while $\rho_{A\downarrow}$ remains below 0.5. This results in a long-term stable spin polarization in A. Figure \ref{fig11}(d) show the spin polarization $P_{As}$ of A. $P_{As}$ shows a high value in the short time, but does not decay to zero in the long time, leading to a stable spin polarization in A. 
These results are identical to protein-like molecules as the chiral molecule bridge B.

These results of D-dsDNA-A system--fully consistent with our protein-based calculations--confirm that our theory captures universal dynamical mechanisms underlying the CISS effect, irrespective of molecular type or HOMO-LUMO gap behavior.

\end{document}